# Comparing the Performance of Traffic Coordination Methods for Advanced Aerial Mobility

Ítalo Romani de Oliveira, Euclides C. Pinto Neto, Thiago Matsumoto, Huafeng Yu,
Emiliano Bartolomé, Guillermo Frontera, Aaron Mayne

Boeing Research & Technology

*Abstract*— Traffic Management in Advanced Aerial Mobility (AAM) inherits many elements of conventional Air Traffic Management (ATM), but brings new complexities and challenges of its own. One of its ways of guaranteeing separation is the use of airborne, stand-alone Detect-And-Avoid, an operational concept where each aircraft decides its avoidance maneuvers independently, observing right-of-way rules and, in specific implementations, some form of pairwise coordination. This is a fundamental safety element for autonomous aircraft but, according to our research, is not sufficient for high-density airspaces as envisioned for urban environments. In these environments, some way of explicit and strategic traffic coordination must be in place, as done for conventional ATM. For efficiency reasons, ATM is evolving to more flexible uses of the airspace, such that the use of dynamically allocated corridors is a rising concept for AAM. These strategic forms of traffic coordination are potentially highly efficient if the aircraft adhere to their trajectory contracts and there are no significant perturbations to the traffic. However, if significant perturbations occur, such as loss of data communication, or the sudden appearance of an intruder, a centralized system may not react appropriately in due time. In busy scenarios, even small deviations from plans may compound so rapidly as to result in large differences in the overall achieved scenario, resulting in congestions and convoluted conflicts. Therefore, it is worth studying traffic coordination techniques that work locally with shorter look-ahead times. To that end, we explore an airborne collaborative method for traffic coordination, which is capable of safely solving conflicts with multiple aircraft, stressing its capabilities throughout a large number of scenarios and comparing its performance with established methods.

*Keywords—Advanced Aerial Mobility, Traffic Coordination, Conflict Detection and Resolution, Flight Efficiency*

## I. INTRODUCTION

The presently widespread concept of Air Traffic Management (ATM) combines airspace structure elements such as sectors, airways and terminal routes, with various surveillance and communication means, and software for control and decision making support. This system is the main reference for the creation of an analogous system that will manage traffic of new forms of aerial mobility, which are within the limits of smaller "universes" constituted by the urban and metropolitan areas. These new concepts of aerial mobility are referred generically as Advanced Aerial Mobility [1] [2], a term which is becoming progressively more popular. In order to better distinguish it from the already established ATM System, we refer to that pre-existing ATM concept as Intercity ATM. Different from that, AAM covers the concepts of Unmanned Traffic Management (UTM), for small Unmanned Aircraft Systems, as well as the new types of passenger-carrying aerial vehicles designed for Urban Air Mobility (UAM).

Well-substantiated Concepts-of-Operations for AAM and UAM [3] [4] count on the use of airspace structures as a fundamental means of order and safety. However, complying with fixed corridors clearly affect the efficiency of individual flights, especially in moments when there is little traffic. This fact has led to the rise of flexible Operation Volumes [5] and Operation Volume Contracts (OVC) [6], as a way of guaranteeing safe separations without the need of fixed airspace structures. Closely related to the airspace structuration or lack thereof is the amount of airspace reserved for each aircraft. The larger the exclusive extent of airspace for safe operation of an aircraft is necessary, the least capacity the enclosing airspace will have and more congestion will happen, causing inefficiency in the time dimension [7].

Another relevant factor influencing airspace capacity and performance is the coordination method used to orchestrate who occupies certain portion of the airspace in certain time interval and the rules governing such orchestration. In certain contexts, this may be as natural as we concede or follow passage in our cities' streets and walkways, but in other contexts, the airspace occupation is highly constrained and subject to strict and centralized coordination. The former context is that of airspaces with low occupation densities, while the latter is more appropriate in densely occupied airspaces. In a recent paper [8], we brought to the light some issues of uncoordinated Conflict Detection and Resolution (CD&R) when they are used in densely occupied airspaces. In this paper, we further investigate those issues using more realistic models, considering both unperturbed and perturbed scenarios, and explore other aspects of this problem.

## II. SIMULATION MODEL

### A. Basic Airspace Structure

Our airspace model is bi-dimensional and based on a lattice structure composed by equilateral triangles, as shown in Fig. 1.

In a UAM context, the vertices in the lattice could represent vertiports, and the edges the airways or air links between them. However, in a more generic context, the vertices can be just waypoints of an en-route airspace, while some of them, especially those at the border, could be vertiports or just entrance-exit waypoints. Departing from the central vertex of the figure and going along the upward link, we can see that link as one of the radials of a hexagon of radius one ($r = 1$) and, joining the second upward link to the segment, we see that segment as a radial of a hexagon of radius two ($r = 2$).

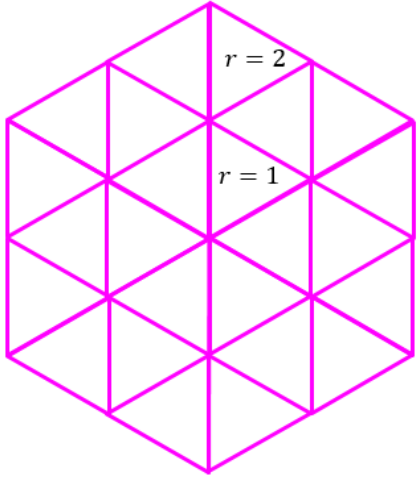

Fig. 1: Basic airspace structure of waypoints and air links.

This hexagonal pattern gives the name Hexagonal Lattice to this structure. It is a planar structure and this implies that the aircraft maneuvers occur at the same altitude. While there are 3D lattices that could be employed, our analysis so far is restricted to the horizontal plane. This restriction is not unlikely in practice, because there may be airspaces constrained above by manned aircraft and below by terrain/obstacles. This basic structure is the same used in [8], but here it is associated with more realistic elements, presented below.

### B. Continuous Trajectory Dynamics and Airspace Cells

An aircraft maneuvers in curves and this feature is highly relevant in conflict resolution maneuvers and in any trajectory-related study in scales small enough to the radius of the curve to become significant. The aircraft model that we selected has fixed wings and a typical curvature radius of 350 m (~1,150 ft.) and this measure is strongly significant to the separations currently considered in UTM and AAM. Rotary wing aircraft are capable of smaller radii or even stopping and reversing course in the air, but the accelerations, decelerations and energy consumption of such maneuvers have to be taken into account.

In order to provide room for maneuvers, we establish cells around the waypoints of Fig. 1 and use these cells as atomic allocation units, each one subject to reservation for exclusively one (or none) aircraft at each moment. In order to partition the airspace in equally sized cells according to the vertices of the triangular lattice, the resulting cells become hexagonal, according to Fig. 2. Grid cells are one of the proposed means for remaining Well-Clear in the study [9]

We number the cells in circular order, in concentric circles, so that, given that single cell number, it is possible to easily know the radius at which it stays (the distance in links to the center) and, besides, to locate it univocally in the array according to a polar-like coordinate system.

The departure and destination points, as well as the waypoints employed in the structured traffic coordination methods herein used, must be the cell centroids. However, the aircraft turning maneuvers can, in principle, use any point in the cell, as long as the aircraft respect a contracted occupation interval of time for that cell.

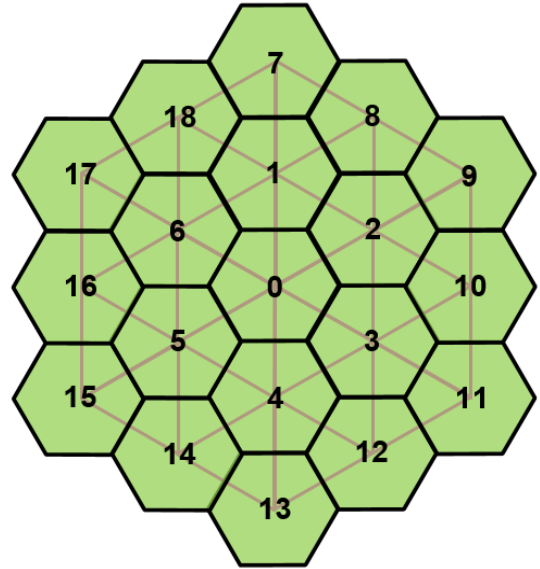

Fig. 2: Airspace partitioned in cells according to the triangular lattice.

An example of traffic scenario in this airspace is given in Fig. 3. The scale of the figure is in meters, from which it is possible to check that the distance between cell centroids is 4,000 m (4km), a value set empirically, but that can be decreased in future research. There are four aircraft in the scenario, identified by the labels V0, V1, V2 and V3 (V for vehicle). The origin points are marked with a circle, and the destination points are marked with 'X'.

In order to facilitate trajectory prediction and keep the airspace algorithms simple, we imposed a constant sector occupation time, when subject to the coordinated algorithms. This way, whichever the entry and exit side of the cell, the occupation time will be constant. We can notice this constraint taking effect in the first and second curves of aircraft V0. The first curve makes a turn of 120 degrees and, in order to compensate for the length of the curve, the maneuver starts much before the cell centroid, such that it ends up on the outer side of the curve. An analogous maneuver is performed in the second curve, but as the turn angle is only 60 degrees, the curve passes closer to the cell centroid.

The simulator checks the distances among the aircraft each two seconds and registers the smallest spacing incurred for each one throughout the scenario, the position at which it occurred, the identity and position of the other aircraft involved. We name this event as the Closest Point of Approach (CPA) and represent it in the Fig. as a black dashed line linking the positions of the two aircraft involved in the event, positions at the CPA represented by small colored stars.

In this figure, aircraft V0 had its CPA with aircraft V1, but aircraft V1 had its CPA with aircraft V3. V3 had its CPA reciprocally with V1 (their black dashed lines are superimposed and thus seem continuous), and V3 was involved in the CPA of aircraft V2. In this figure, the trajectories are strategically deconflicted and noticeably the CPAs occur when they are at the borders of a cell.

We can also notice that V0 makes a detour in the beginning and wonder why that happened. The reason for that is that it has the same destination and same distance to destination as of V2, being chosen to be the second to arrive, thereby avoiding to arrive at the same time.

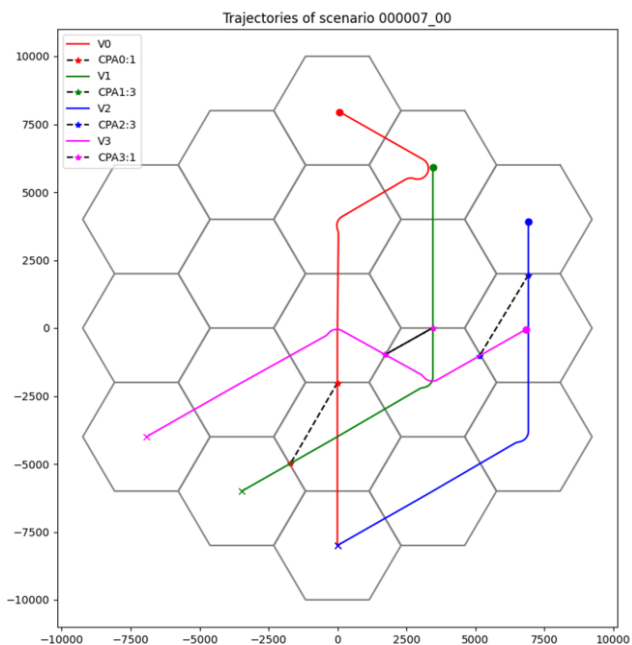

Fig. 3: Aircraft trajectories from an example traffic scenario.

We implemented this airspace model as client to the Skyway Simlite simulator [10] and chose as aircraft model the so-called Avión Ligero de Observación (ALO) [11], a mid-range unmanned aerial platform from the Spanish INTA with maximum take-off weight of 60 kg and cruise speed of 62 knots. The simulations set the speed to around 86 knots, still within the operating range of the aircraft. The distance between cell centroids of 4,000 m was chosen as to allow enough room for maneuvering inside the cell, so that, even if the aircraft has to make a turn of 180 degrees, the maneuver is fully contained inside the cell, with still some buffer, and the cell occupation time remains constant. This measure makes that the minimum distance between the centroid and the cell border be 2,000 m or 1.08 nautical miles (nmi). The simulations have shown that, with proper trajectory control, the strategic and collaborative airspace algorithms from [8] are effective to keep traffic separation for such aircraft type.

In the next section, we revisit the methods used for Traffic Management and / or Conflict Detection & Resolution employed in the scenarios.

## III. CONFLICT DETECTION & RESOLUTION METHODS

Conflict Detection & Resolution (CD&R) is an important aspect of Traffic Management. Traffic conflicts occur when a vehicle agent has a trajectory intent that becomes unsafely close to another vehicle's trajectory intent, such that the risk of collision rises and becomes unacceptable. If the trajectory intents remain unchanged and the vehicles follow these trajectories, the potential conflict becomes an actual conflict or, in other words, a Loss of Separation (LoS). A LoS situation has an unacceptably high risk of resulting in a collision, but that not necessarily happens.

In any case, these general concepts of separation and conflict need mathematical and numeric definitions, which we will provide below. In this section, we describe the methods used in high level without referring to numeric values of the parameters used.

### A. Detect-And-Avoid (DAA)

In this method, each aircraft keeps separation by its own means. It may have some form of implicit or explicit coordination among the aircraft, and should be effective in cases where one or more aircraft are uncooperative, however these cases are out of the scope of this study. The fundamental feature that we identify in this method, in relation to the others, is that it is designed so as that a first aircraft, called ownship, is capable of, entirely by its own means, detecting conflict with a second aircraft, called intruder, and, as consequence, performing a maneuver that will preserve the separation between them, whether or not the intruder is operating a similar DAA capability. For the cases where a third or more aircraft are involved in the conflict, the method implementation may have a higher-level logic that prioritizes the intruders and maneuvers in order to choose the least risky action, however the degree of effectiveness of such prioritization has not been clearly demonstrated yet.

DAA reached industry-standard maturity level in 2017 and is in its second revision, named DO-365B [12], although this standard document assumes that the aircraft are remotely piloted. The implementations of it that are most accepted by the industry are DAIDALUS [13] [14] and ACAS-Xu [15]. This research adopted DAIDALUS in our simulation environment because its source code was publicly available by the time we started these studies.

The DAIDALUS DAA algorithm provides alerts of conflicts according to the severity levels established in [12] and directions to which the ownship must maneuver in order to regain DAA Well-Clear state (DWC), that is, to solve the conflict and avoid LoS. However, the exact trajectory which the aircraft will follow during and after deviation is up to the aircraft mission and control systems. After DWC is regained, the aircraft will have to choose a path to its destination and do a resume maneuver to follow it. The resume logic is not part of the DAA standard and none of the cited implementations, so our integrated environment has to supply that logic. Considering these aspects, our DAA logic can be summarized according to the algorithm of Fig. **4**.

| DAA Logic | |
|---|---|
| 1 | Update DAA state with all aircraft's positions and velocities; |
| 2 | If a DAA alert is found: |
| 3 | If there is DAA recovery band clockwise, start or continue turn clockwise; |
| 4 | Else |
| 5 | If there is DAA recovery band counter-clockwise, start or continue turn counter-clockwise; |
| 6 | Else continue on the current path. |
| 7 | Give X seconds for performing the chosen maneuver and go back to line 1. |
| 8 | Else If DAA alerts have been cleared: |
| | Start maneuver to resume path to destination; |
| | Go back to line 1. |

Fig. 4: Summary of the DAA logic in each aircraft from our simulated environment.

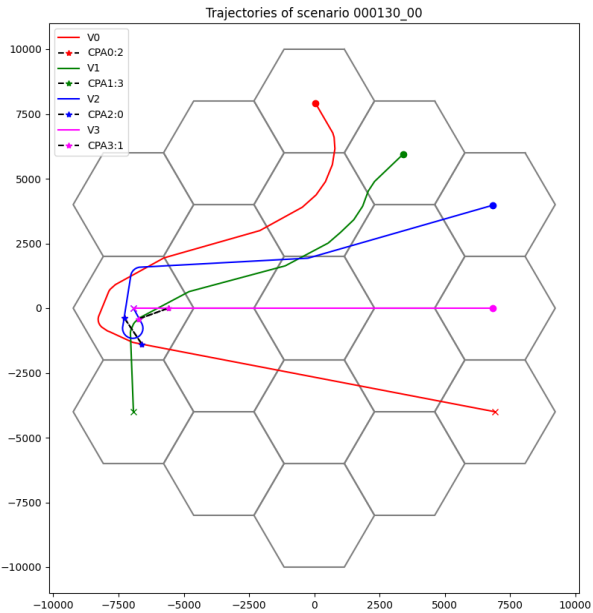

Fig. 5: Example of traffic scenario using DAA CD&R.

This logic, when applied to various different traffic configurations, result in the most varied and interesting trajectories. One example of such combined trajectories is shown in Fig. 5.

Unlike Fig. 3, here the aircraft do not respect cell allocation. Instead, DAA tries to maintain a protection cylinder centered on the aircraft at all times. As the model is bi-dimensional, this cylinder becomes a circle. The radius recommended by the standard [12] is 4,000 ft. (~ 0.66 nmi or 1,220 m), which is respected in the figure's scenario. This value is referred to as Horizontal Miss Distance (HMD).

In Fig. 5, all the aircraft start (from the respective small circle positions) with their ideal courses, pointing direct to destination. Then aircraft V0 (red) detects a conflict with V2 (blue) and steers clockwise. At this moment, there is no conflict between V0 and V1 (green) because V0 would pass behind V1.

However, with the change in V0's direction, a conflict arises between the last two and V1 is the first to detect it. The DAIDALUS logic prohibits it to turn clockwise, so it starts turning counter-clockwise and becomes engaged in a transitive conflict avoidance, V1 avoiding V0, and V0 avoiding V2. This continues for a while until V1 detects a conflict with V3 (magenta) and then all the aircraft become involved in the conflict. Near the time that V2 enters in the central cell, it has passed behind V1 and is already clear from it; however, it detects a conflict with V3 and steers clockwise to avoid it. Between that moment and when V3 reaches its destination, all the other aircraft are avoiding V3, with V0 doing so indirectly because of V1. A little before V3 reaches destination, it has its CPA with V1, which is already ahead of it, and at V1's CPA point, we can see that it starts the final turn to the destination. This move clears angles for V0, which also turns counter-clockwise to seek destination. However, V0 is still not completely clear from V1 and keeps an avoidance direction for a while. A little after, V3 reaches destination at the left most cell on the equator of the figure. At this moment, V2 (blue) is almost to the north of it on the y-coordinate, near 1,600, and V0 (red) is to the southwest of V3 on the same cell. With V3's exit, we can see both V2 and V0 doing sharp clockwise turns, and despite V2 being very close to its destination, it still cannot head directly to it because it still has a conflict with V1, ahead of it to the south. In addition, V0 still cannot head directly to its destination because it now detects a conflict with V2 and has to pass ahead of it to become clear. When this happens, we see the occurrence of the mutual CPA between V0 and V2, thus V0 becomes clear to destination, which is still far ahead on southeast. Shortly after, V1 reaches destination on southwest and V2 becomes clear to destination, which is close behind it. V2 then does a small fishhook maneuver and reaches destination.

In this scenario, V0 (red) was the last one to reach destination and was heavily penalized by the avoidance maneuvers. Its total flying time is 672 seconds at a groundspeed of 44.4 meters per second (~86 knots). If it knew ahead of time the trajectories of the other aircraft, and that the DAA logic would result in such long detour, it could have started the scenario heading eastwards and avoided all the traffic by a very small cost. However, not even with a look-ahead time of 110 seconds, which is the maximum recommended by [12], DAA alone could have anticipated such inefficiencies. During development, we also tried to give preference to "pass behind" maneuvers, whereby V0 would have steered counter-clockwise on the first maneuver and have been cleared much earlier, however this feature degraded the algorithm's effectiveness in the overall set of scenarios, increasing significantly the occurrences of LoS with some Near Mid-Air Collisions (NMAC) [12].

We also learned that DAA has a better performance when the maneuvers of all participants are strongly biased to one rotation sense, either clockwise or counter-clockwise. This is according to right-of-way rules and DAIDALUS tries to observe it, however we tweaked our implementation, so that the logic we used is a bit more biased to the clockwise sense than DAIDALUS'. This rule allowed better performance and was empirically discovered.

### B. Strategic Airspace Allocation

This method aims at decreasing the inefficiencies noted in the previous section and, at the same time, keeping the aircraft duly separated from each other. A central agent receives the flight missions or intents from the aircraft and creates a plan, which determines which 4-dimensional trajectory each aircraft will fly in order to fulfill its mission safely. Usually, this central agent is ground-based and needs to compute trajectories before the aircraft enter the scenario (works off-line). Furthermore, it needs the aircraft to comply with the trajectory within a certain time window. This form of CD&R has high potential of effectiveness in certain AAM contexts, such that there is a considerable amount of work devoted to how to perform it [6] [16] [17] [18]. This is not an entirely new development, since Intercity ATM is analogous to it and has been evolving to 4D Trajectory Management [19] since a number of years ago.

In our simulated environment, the basic airspace allocation units are the hexagonal cells, and we used a Mixed-Integer Programming (MIP) algorithm to provide optimized cell allocation plans that allow only one aircraft per cell at a time, and makes all aircraft to move from origin to destination using the minimum collective amount of time, that is, minimizes the sum of individual flight times.

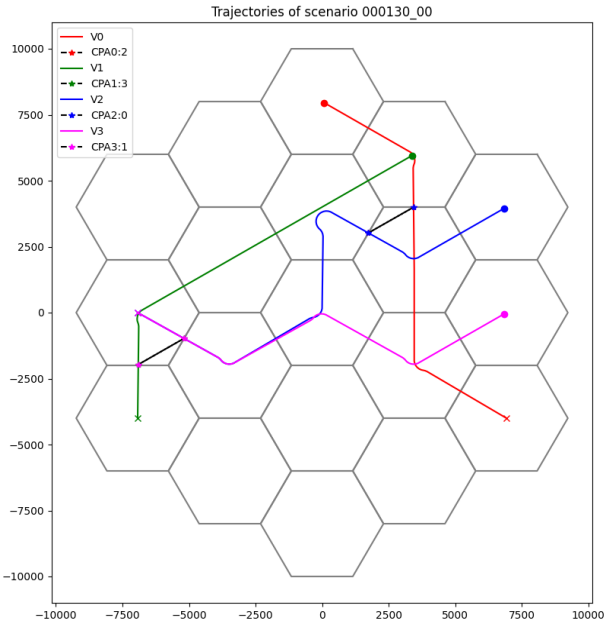

Fig. 6: Example of traffic scenario using Strategic Airspace Allocation.

As typical of a strategic method, the aircraft maneuvers are pre-defined for the entire duration of the scenario. In public operations, the time spans can be considerably large, or even undetermined, until it happens that all the aircraft are stopped, if that ever happens. Existing solutions of this type for Intercity ATM simply truncate the operation time window and periodically recalculate the scenario with actual data, thus requiring a lower layer of continual tactical resolution to ensure separation and thus composing a hybrid solution. In our simulation model, we can cover the entire operation window because both the number of flights and the flight durations are small.

When used to accomplish the same set of flight missions accomplished in the scenario of Fig. 5, the strategic airspace allocation method comes up with the trajectories of Fig. 6. In this solution, all the trajectories must pass through the centroid of every cell they visit, in order to keep separation and maintain constant the cell traversal time, which is a way of simplifying computations. It is not the only solution possible with the strategic algorithm, but the one that resulted from the specific algorithm's implementation. In this result, V0 (red), V1 (green) and V3 (magenta) obtain the shortest possible paths, visiting 3 cells except origin and destination, and V2 is penalized with an extra leg because it has the same destination as V3. As in Fig. 3, the CPAs occur when the aircraft are at the cell borders. The maximum flying time in this scenario is 447 seconds, 33.5 % less than the DAA solution, flying at the same speed, even with its various maneuvers.

### C. Collaborative Airspace Allocation

This method blends elements of the two previous one, in the sense that it does not require that every aircraft have its full path fixed since the beginning, but requires that everyone agrees on the priorities for everyone, and that they explicitly coordinate the maneuvers with the nearby peers, following a certain airspace allocation protocol. This way, the conflict resolution algorithm does not need to cover the entire 4D state-space, thereby being computationally less complex, allowing it to be executed individually by each aircraft and also providing resiliency to perturbations.

One of the key features of this method is that, at certain moments of the evolution of a traffic scenario, near-term airspace resources (cells and edges) are allocated via negotiation between the peers, from which the aircraft draw short-term trajectory contracts that they have to follow until a new negotiation round is needed. This negotiation process is described in [20] and analyzed in [8], but in these studies the negotiation rounds occurred at fixed intervals and the consequent aircraft's maneuvers were fully synchronized. Here, the negotiation is triggered each time an aircraft needs to allocate the next cell in its path and, as in the previous references, an aircraft is not allowed to allocate a cell which is not a neighbor of the cell that it presently occupies. In addition, the aircraft's cell entry and exit times are not synchronized and can occur at any moment in a finite time.

In order to simplify the aircraft's trajectory planning and prediction algorithm, they adopt a maneuvering rule whereby the cell occupation is constant for a certain speed, similarly to what is done for the Strategic Airspace Allocation method from Section III.B.

When running this method for a mission scenario similar to the previous examples, we obtain the trajectories of Fig. 7. As it can be noted, here the aircraft also have to pass at or near the cell centroids to better manage separation and time. We can also note that all aircraft except V3 accomplish the mission with paths of minimum length. In this scenario, the aircraft's priorities in conflict resolution follow the numerical order of their id's, thus V3 has the least priority and is the first to be diverted from its optimal path in case of conflict. Given that it was assigned two extra legs of deviation, it is the last one to accomplish its mission, which happens after 516 seconds of flight.

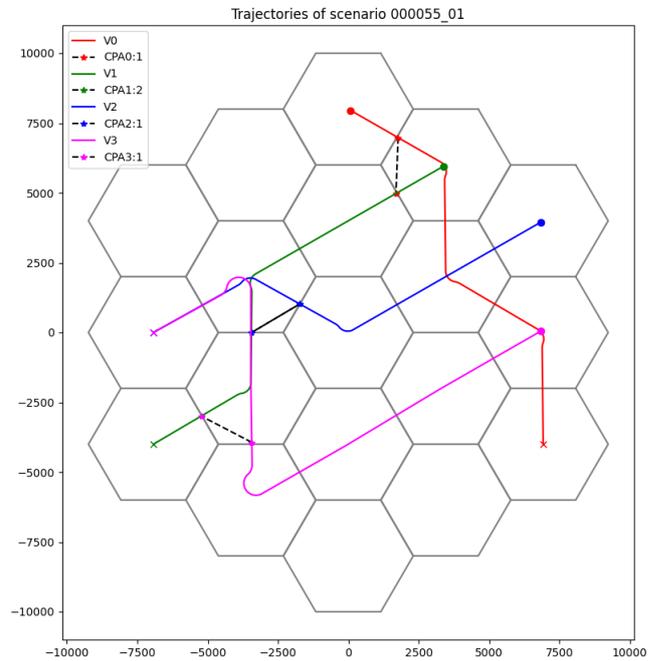

Fig. 7: Example of traffic scenario using Collaborative Airspace Allocation.

The key driver in proposing and using this alternative CD&R method is to provide more redundancy to conflict resolution in mid-density and high-density airspaces, given that we identified shortcomings of DAA in solving conflicts with multiple aircraft. Furthermore, strategic deconfliction requires huge efforts in Verification & Validation and costly infrastructure investment, while this method is based on simpler logic and does not require infrastructure.

IV. SIMULATION SETS

We ran millions of scenario instantiations to statistically compare the performance of the three methods above described. Each of the method had its scenarios configured in certain ways to allow observation of the most relevant aspects and with the proper reference parameters, including the occurrence of perturbations to the Strategic Airspace Allocation method. In these scenarios, there are four aircraft and each one has its mission defined by an origin and a destination vertex, both of which must be on the outer airspace cells, each aircraft starting in a separate cell, and the minimum path between origin and destination cells must have at least three intermediate cells. This is a way of excluding short trajectories and generating more conflicts per scenario. Another feature of such choice is that an aircraft will always cross either the equator or the meridian of the flattened airspace.

Given these conditions, we generate all the possible combinations of aircraft initial positions and destinations, but applying such combinations in a fixed aircraft ordering, such that an aircraft with a certain id will receive certain origin-destination combinations that other aircraft id will not receive, and vice-versa. In order to fully explore the vast number of scenario possibilities, one should consider all permutations of aircraft ids, because either the CD&R algorithm or the simulation implementation, executed on a sequential CPU, uses the aircraft id as a criterion to execute some state change of a certain aircraft before another in a deterministic way, thus eliminating certain scenario possibilities. An analysis closer to such full-combinatorial coverage was performed in [20] for Collaborative Airspace Allocation, yet with a simpler simulation model and still some uncovered cases. Here, the fixed aircraft ordering was deemed enough to generate a satisfactory coverage of the space of scenarios, given the computation times.

Below follows the description of each set of scenarios.

*A. DAA Unperturbed*

In this scenario set, all aircraft start at the same time at their respective origin cell centroid, heading straight to their respective destinations. These are pure DAA scenarios as described in Section III.A, using DAIDALUS as the DAA algorithm exclusively on the horizontal plane, with base configuration according to DO-365B [12], and aircraft turning capability adjusted to the chosen aircraft model, which has a typical turning rate of 6.5 degrees per second.

In order to define the aircraft's protection volume, DAIDALUS uses a parameter named DTHR (Distance Threshold), which corresponds to DO-365B's HMD (Horizontal Miss Distance) and DMOD (Distance of the Modified Tau) and, in a succinct definition, corresponds to the horizontal radius of a protection cylinder. We started using the value DTHR=4,000 feet (0.66 nmi) recommended by Table 2-24 of DO-365B, however in the simulations we noted a considerable number of violations of this protection radius, so we performed another simulation set using DTHR=1.0 nmi. In the comparison of the results below, these scenario configurations are referred to as `daa_u0.66nmi` and `daa_u1.00nmi`.

Considering the above explained methodology to generate combinations of traffic configurations, a total of 122,415 variations of 4-aircraft origin and destination points was simulated for each or the two DTHR configuration variants.

*B. Strategic Unperturbed*

Here, the same 122,415 traffic configurations generated for the previous set of scenarios were simulated, but using Strategic Airspace Allocation as described in Section III.B, with DAA switched off. The initial heading of each aircraft is according to the path provided by Strategic Allocation. This scenario is referred too simply as `strategic_u` in the comparison below.

*C. DAA Recovery*

In this scenario set, three aircraft start according to Strategic Airspace Allocation at time $t = 0$ and, after that, at time $t = 30$ seconds, a fourth aircraft appears in the scenario as an intruder, at one of the outer cells' centroids. The idea is to represent a disruption to the strategic plans, which may happen for various reasons, including, but not limited to, the following contingencies:

- The intruder aircraft inadvertently enters strategically allocated airspace as if it were in a DAA-only airspace;

- There's a major outage (due to hardware failure, cyber-attack or bug) of the central system of Strategic Airspace Allocation and operations should continue;

- The intruder aircraft has a technical failure, which turns it unable to communicate with the central traffic coordinator and/or fly on strategic mode, and cannot compute a safe emergency landing on an alternate destination.

When the intruder enters the scenario, it heads directly to its destination, and we initially considered both the cases where the intruder is DAA-uncooperative and DAA-cooperative. The uncooperative ones represent the situations where the intruder either has unquestionable higher priority, is illegally not willing to cooperate, unable to cooperate because of some technical problem, or simply unaware of the exceptionality of its uncooperative condition. After some rounds of simulation, it became clear that such uncooperative scenarios presented worse performance so we deemed unfair to compare it with the other scenarios with the goal of evaluating the potential of each method. We might consider that the cases of top priority missions, such as medevac or law enforcement, could receive a special treatment from the central traffic manager (if present), fly at another altitude, etc., so as to minimize collision risk. Thus, here we present only the cases where the intruder is DAA-cooperative.

We assumed that, once the intruder enters the scenario, the other aircraft will continue on their strategically designated paths, but with DAA activated, such that, if DAA detects a conflict with any other aircraft, it will perform the DAA-indicated avoidance maneuver. Once DAA is cleared, it uses certain criteria to define the resume path:

1. Check if the current cell is in the continuation of the strategically designated path and, if it is, continue on the strategic path from that cell on (it may require some direction adjustment to align with the path).

2. Else, if a neighbor of the current cell is in the continuation of the strategic path and is in a favorable position (does not require "going back" on the path), start a maneuver to seek that cell's centroid and continue on the strategic path from that cell on;

3. Else, start a maneuver to seek the destination cell's centroid and try to proceed directly to that cell.

In reality, an aircraft that is far from the intruder could not detect it by DAA and continue to assume that the aircraft which are closer to itself are still following strategic guidance. To be more concrete, let us suppose that the intruder V3 is out of DAA sensing range from the regular participant V1. If the regular aircraft V2 starts a DAA maneuver to avoid V3 and becomes closer to V1, this one may continue to assume that V2 had this maneuver strategically planned and ignore early DAA alerts. To rule out such cases, we assume that:

1. DAA is always active but is ignored if and only if the central traffic coordination system is active and all aircraft are verifiably following strategic guidance;

2. If there is a major failure in the central traffic coordination, all aircraft will be able to detect this situation and start giving top priority to DAA;

3. If an intruder enters the scenario, any aircraft that is in DAA range to it will know that it is an intruder and give priority to DAA alerts;

4. If any of the "regular" participants deviate from its strategically defined path, for any reason, all the aircraft which are able to sense its position will know that it is deviating and follow DAA guidance, if DAA alerts occur;

5. If the central traffic coordination system is correctly functioning and able to detect an intruder or any deviation from the strategic plans, it will emit alerts to all participants that may be affected by such exception.

As a reasonable protective measure, we consider that, even when the intruder is approaching the cellular airspace from outside, the regular participants' DAA that are in sensing range to it can detect it and start deviation maneuvers if DAA alerts occur.

Given these assumptions and definitions, we generated and simulated all the possible traffic configurations, resulting in a higher number than for the previous sets. As the difference between an intruder and a regular member is not only the aircraft identification number, there is one more combinatorial factor (intruder / non-intruder), resulting in a total number of 373,680 traffic configurations.

In addition, for the same reasons explained in Section III.A, we present the analysis of two variants of this scenario: the so-called `daa_rec0.66nmi`, with protection radius DTHR = 0.66 nmi (4,000 feet), and `daa_rec1.00nmi`, with protection radius DTHR = 1.0 nmi.

### D. Collaborative Recovery

This scenario set uses the same 373,680 traffic configurations of the previous section for the same situations of perturbation to the Strategic Airspace Allocation with an intruder, however DAA is switched off and, instead, the ensuing conflicts are resolved by means of the Collaborative Airspace Allocation method described in Section III.C.

As in the previous section, the non-strategic intruder is cooperative and certain safety condition is imposed before it suddenly appears in the airspace. The condition is that it waits outside the airspace (e.g. on the ground, on a vertiport, in a neighboring airspace, etc.) if the first cell that it needs to allocate is already allocated, that is, if there is a conflict in the time of entrance. As a simple rule, we defined that the intruder will wait 40 seconds in those situations, after which it will perform a new attempt to enter, and so on. The name given to this scenario set is `collab_rec`.

### V. STATISTICAL PERFORMANCE ANALYSIS

We ran the simulation sets described in the previous section and collected some performance metrics for each of them, according to the following description.

#### A. Actual Horizontal Miss Distance (HMD)

The DAA standard DO-365B [12] establishes the Horizontal Miss Distance (HMD) as the safe separation threshold that must be respected among the aircraft at all times. However, its mandatory test cases include only scenarios with two aircraft and, even that, they do not cover the whole universe of two-aircraft encounters. It sounds reasonable to accept, as we found, that in a large variety of scenarios with 4 aircraft, each one following a closed-loop guidance to accomplish their missions, there occur cases where the distance between two aircraft reach a value smaller than the required HMD, even when using algorithms that satisfy the DO-365B criteria. That is why we denominate this metric as "actual" HMD, in order to distinguish it from the required HMD. It is defined here as the minimum distance between two aircraft that occurred throughout a scenario instantiation.

Fig. **8** shows the mean and minimum values of HMD for each scenario type, with an indication of the required HMD of 0.66 nmi.

#### B. Rates of occurrence of unsafe events

Looking at mean and minimum actual HMDs is not the only way to analyze unsafe events in the scenarios. It is also important to know the rate of occurrence of HMDs, which gives an indication of its probability in the universe of scenarios, and other unsafe events that were observed in the simulations, defined below.

*a) HMD Violation:* such event happens when the actual HMD is below the threshold of 0.66 nmi; in our context, this is the equivalent of a Loss of Separation (LoS).

*b) Airspace Excursion:* it happens when some aircraft goes out of the cellular airspace as a result of a conflict resolution maneuver. We consider this as an unsafe event because there may be other aircraft or obstacles in the airspace surroundings. The surrounding airspace may be reserved for other operations and as so should not be invaded. In fact, to evaluate this metric we added a buffer and considered that an excursion happens when an aircraft becomes at a distance from the central cell centroid of more than 2.6 times the distance between cell centroids (10,400 m or 5.62 nmi), a circle which covers all cells and a little more.

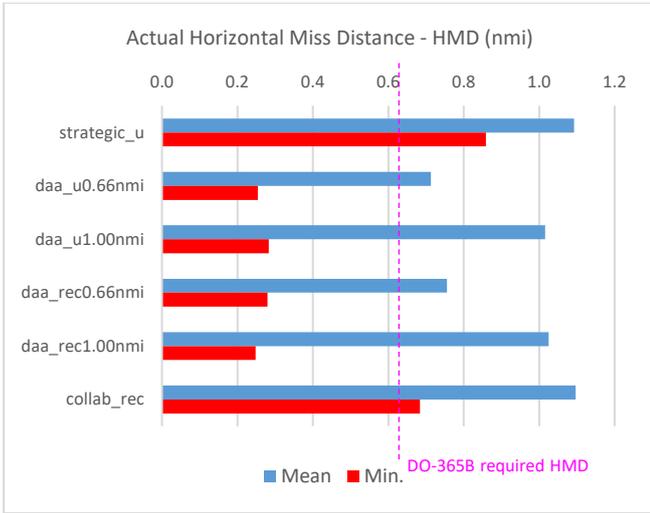

Fig. 8: Actual Horizontal Miss Distance comparison.

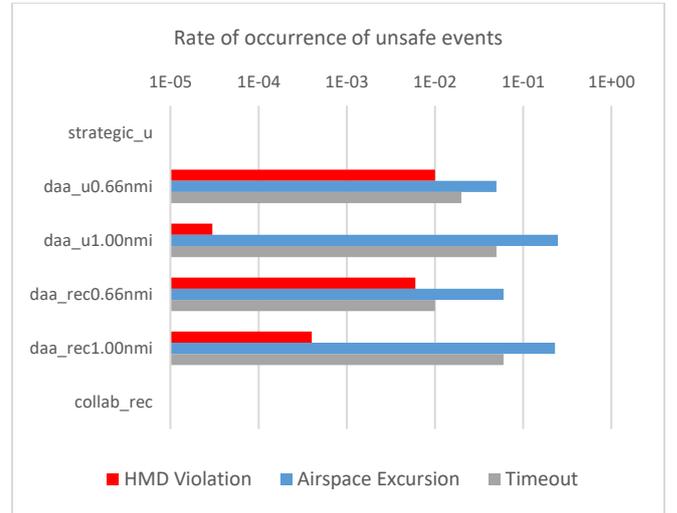

Fig. 9: Rates of occurrence of unsafe events.

*c) Timeout:* this is an issue emerging from the lack of coordinated prioritization among conflicting aircraft when they need to resume to their respective mission paths, already observed in *[20]* and *[8]*. Imagine that aircraft V1 and V2 are involved a conflict and both perform maneuvers to avoid each other. Once any of them reaches a state where it no longer detects conflict, it will try to change course and resume to the mission that it has to accomplish. The resume reference can be the next waypoint in the path or the destination point. It may happen that this resume maneuver generates a new conflict and both of the aircraft have to either perform a new deviation or continue the ongoing deviation. This situation can repeat itself indefinitely until some aircraft finds itself in fuel emergency. When this occurs in a simulation where fuel is not accounted, it can have no end, so we established the timeout limit as 1,000 seconds of simulated time. We provide more details about timeouts in Appendix A.

The rates of occurrence of the events above was computed on the assumption that each traffic configuration is equally likely to occur. In practice, the probability of each traffic configuration will depend on the frequency and times of occurrence of the various flight missions, which may be highly unbalanced. The resulting rates are shown in Fig. **9**.

### C. Efficiency and Equity

In small and uniform airspaces as here modeled, the optimum path for a single aircraft between two points is a straight line. When we add more aircraft to the airspace, conflict resolution maneuvers are necessary and one or more aircraft will deviate from its optimum path and fly extra distances in relation to their unimpeded optimal paths. We measure the efficiency of a CD&R method by computing the mean value of such extra distance flown in a scenario instantiation, among the participant aircraft, and then taking the mean value among all scenario instantiation for a given method.

However, looking only at the mean value hides the fact that one aircraft is being inequitably treated and taking too much of the deviation costs in relation to its peers. Thus, in order to check the equitability property of each method, we also compute the sample standard deviation among the four aircraft in each scenario, and then take its mean value among all scenario instantiations in each set. The results of such evaluations are shown in Fig. **10**.

### D. Comparative analysis

After analyzing the results of the preceding sections, these are the results that we would like to emphasize:

*a)* The Stragic Airspace Allocation method, when unperturbed (`strategic_u`), is the the safest and most equitable of the methods. It is the second most efficient, losing in mean extra distance only to the unperturbed DAA with DTHR=0.66 nmi (`daa_u0.66nmi`).

*b)* Considering the high rates of unsafe events of `daa_u0.66nmi`, we can say that it is out of consideraraion as sole means for CD&R in these scenarios with 4 aircraft, as here modeled and implemented. Thus, its efficiency is overshadowed by its lack of safety, and `strategic_u` becomes the most efficient one among the feasible solutions.

*c)* `daa_u1.00nmi` might be acceptable as primary means of separation, when looking only to its HMD violation rate, however there should be additional means to avoid airspace excursions and timeouts. Even so, its inefficiency is high and its equitability is the worst one.

*d)* Both DAA-recovery scenarios have rates of unsafe timeouts and excursions as high as the unperturbed DAA scenarios, thus needing extra mechanisms to decrease those rates. Regarding only HMD violations, `daa_rec1.00nmi` might be acceptable.

*e)* The collaborative-recovery scenario `collab_rec` has no timeouts and no airspace excursions, due to the fact that has explicit coordination among the aircraft. It has also no HMD violations, thus making it acceptable for the tested set of scenarios.

With these observations in mind, it seems unquestionable that Strategic Airspace Allocation has the best performance and should be used whenever practical as the default means of separation in dense airspaces. However, Collaborative Airspace Allocation is equally free of unsafe events, albeit with smaller miss distances, due to the late appearance of an intruder in its scenarios, a situation to which it demonstrated resilience.

It seems worth to take a closer look in the actual miss distances of the DAA-unperturbed scenario `daa_u1.00nmi`, as its violations occur with low frequency and could be fixable. In fact, Fig. **11** seems to have an interesting indication.

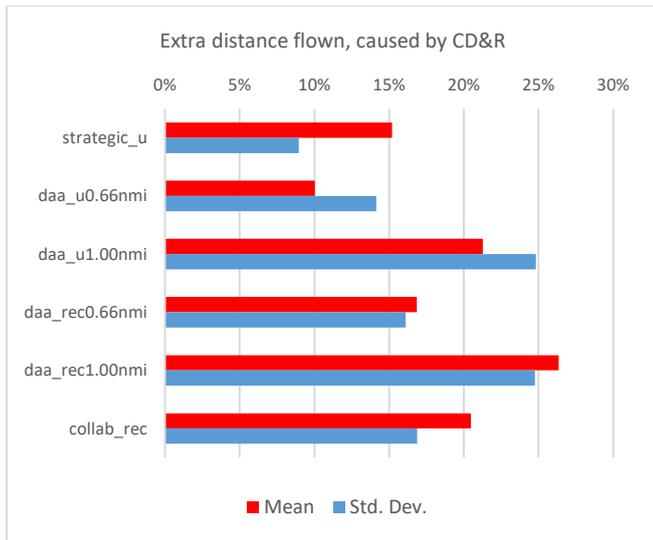

Fig. 10: Extra distance mean and standard deviation.

It shows the probability distribution, or equivalently, the histogram of the actual horizontal miss distances found in the 122,415 traffic configurations of the `daa_u1.00nmi` scenario set. It is possible to observe there just a few occurrences of HMD < 0.66 nmi on the left. In the sample, they represent only 4 distinct instances. Then, looking at the trajectories of each of these instantiations, we found that all of them have convoluted multi-aircraft resolution maneuvers associated to timeouts. Therefore, the solution for the timeouts will probably fix this scenario set almost entirely, remaining only the airspace excursions to be solved.

Actually, if we consider the standard ASTM F3442/F3442M – 20 [21], only one loss of separation occurred in this sample, with the value of 1725 feet, smaller than the 2,000 feet required by the standard.

The development of simple and reliable solutions to eliminate timeouts is included in the scope of future works. This seems to require some form of explicit coordination associated to a mechanism to guarantee full priority ordering among the aircraft. Such mechanism may require some randomness, which would make verification harder. More details about the timeout events are in Appendix A.

## VI. FINAL REMARKS

We compared three distinct methods of CD&R in a dense airspace, simulating about 1,5 million scenarios just for the results shown in this paper, plus a couple of million others which were part of the experimentation process. We confirmed most of the findings from [8] and some from [20] with a more realistic aircraft simulation model. We went beyond that and analyzed the performance of recovery scenarios for the Strategic Allocation Method, which is the best performer when unperturbed, however the less flexible to perturbations. The Collaborative Airspace Allocation method has enough performance to be an alternative to Strategic Allocation, with lesser efficiency, but with more resilience and simplicity. Our findings lead us to think that DAA, as currently defined by DO-365B, does not have performance to be used as the primary means of separation in densely occupied airspaces by unmanned aircraft, despite being still necessary due to its unique capability of avoiding uncooperative intruders. We have not considered the presence of manned aircraft amongst the UAS, but that is likely to pose even more challenging situations to the CD&R methods.

Among the next steps that we intend to give in this research are the inclusion of vertical and speed-control maneuvers; assessments on which conditions would be necessary to decrease the airspace cell sizes, including the use of ACAS-sXu definitions [22].

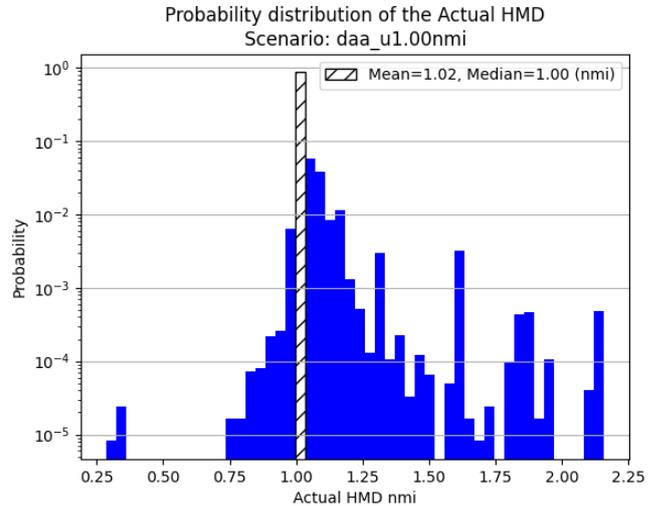

Fig. 11: Probability distribution of the Actual Horizontal Miss distances of the scenario DAA-unperturbed `daa_u1.00nmi.`

APPENDIX: TIMEOUT MANEUVER CHAINS

For the same reason that it does not exist a universal procedure to demonstrate that a computer program terminates, it may be impossible to demonstrate that a certain multi-aircraft scenario, using certain CD&R method, will allow all aircraft to accomplish their missions, if there is no fuel limitation involved. The combined trajectory space can be infinitely huge and impossible to fully cover. Also, modestly complex program logic may be intractable by formal verification methods. These aspects of a dynamic traffic scenario are far beyond the scope of existing DAA standards and UTM /AAM Concepts of Operations.

If there is fuel limitation involved, several conflict resolution maneuvers can occur in succession, repeated or not, until some aircraft becomes low on fuel and has to begin an early termination trajectory, which often leads to an alternate destination. That could occur in reality. In the simulated world, if there is no fuel nor time limitation, the maneuver successions could extend to infinity. Thus, unless one has certainty that the simulated scenarios will always have all aircrafts' missions accomplished timeouts or finite fuel loads must be observed and properly handled.

Despite we believe that the aircraft agent's mission planning and execution logic that we developed can still be improved to better co-exist with the DAA logic, their present interaction still generates some undesired effects and leads to timeouts preceded by the most intriguing chains of maneuvers.

We found two basic types of timing out chains. The first one is linear and identifiable by two or more aircraft flying far out the cellular airspace, such as that of Fig. **12**. In this scenario, aircraft V1 (green) has the same destination as V0 (red), which finished at the southmost cell. V1 could not take a resume path because was in conflict with V3 (magenta), which had its destination the same as V2 (blue). On its turn, V3, despite having passed very close to its destination, could not turn towards it because was in conflict with V1. The conflict between V1 and V3 extended linearly until timeout.

Another type or timeout chain has loops of various shapes, such as that of Fig. **13**.

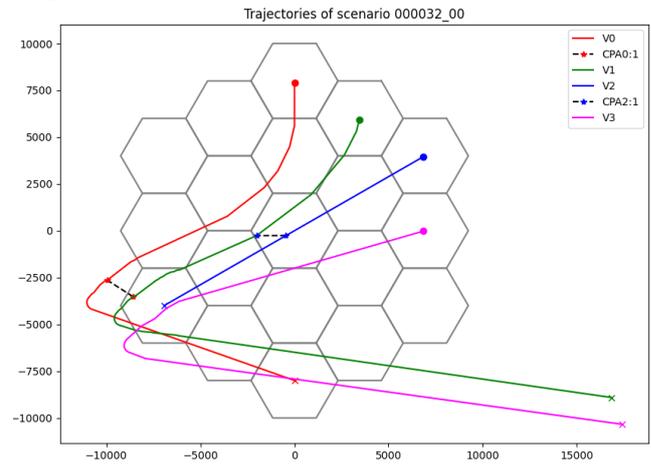

Fig. 12: Linear-type timeout chain of maneuvers.

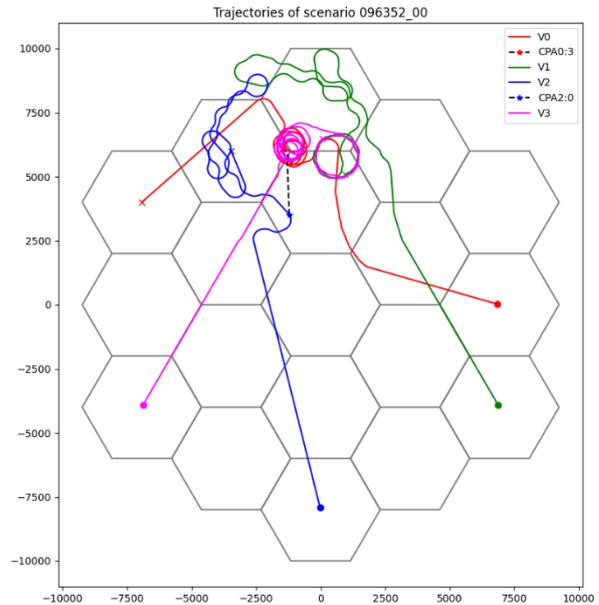

Fig. 13: Loop-type timeout chain of maneuvers.

This is one of the four scenarios of the simulation `set daa_u1.0nmi` with HMD violation, pointed out in the explanation of Fig. 11. In this scenario, all aircraft become involved in the conflict, but only V0 (red) and V1 (blue) can reach their destinations before timeout.

There are many more intriguing examples of timeout maneuver chains, however they will be investigated in future research. One of the next steps in this research is to devise a reliable method to eliminate such situations without depending on human intervention.